# Privacy Shielding against Mass Surveillance

Kashyap.V[#1], Boominathan.P[#2]

[#1]*B Tech CSE, School of Computing Science and Engineering (SCSE), VIT UNIVERSITY, VELLORE*
[#2]*Assistant Professor (Sr) School of Computing Science and Engineering (SCSE), VIT UNIVERSITY, VELLORE*

*Abstract*— Privacy Shielding against Mass Surveillance (PSMS) provides a step by step tactical approach to protecting the privacy of all the users of the internet from mass surveillance programs by the governments and other state agencies. Protection of one's privacy is of prime importance and Privacy Shielding provides the right means against mass surveillance programs and from malicious users trying to gain access to your systems. Although protection is difficult when massive government agencies like the NSA and GCHQ target internet users for surveillance, it is possible because the target is not 'you' as an individual but the entire mass as a whole. With the right approach and a broad perspective of the term 'Privacy', it is possible for one to freely access and share information over the internet without being victims of surveillance.

*Keywords*- Anonymity, Privacy, Protection, Surveillance.

I. INTRODUCTION

Surveillance and censorship has become the most pressing concern to a citizen's privacy and this calls for wide spread awareness among the users of the internet [1]. With over 360,985,492 internet users world-wide out of a population of 7,017,846,922[2] and the recent advancements in the Internet of Things, the issue of privacy cannot be neglected any longer. The Internet of Things, IoT for short is the ideal that every object in your life is web-connected and communicating with each other. This home network can include items as mundane as a coffee mug, smart appliances such as a washing machine, heating and lighting controls, or even sensors attached to your children and pets [3]. About 50 billion machines and devices could be linked by 2020, according to Cisco Systems, a leader in the IoT movement. Such smart devices are already being used, for example, to check soil moisture in vineyards, control the carbon emission of factories, alert drivers to traffic jams, and monitor patients' blood pressure—all without human intervention [4]. In view of the revelations of mass surveillance operations being carried out by various governments in the world, it bestows a sense of responsibility among us to work towards some serious reforms in our policies. Meanwhile, the only order of the day would be by following Privacy Shielding Techniques that are mentioned in this PSMS. A step-by-step approach is followed to achieve anonymity at all levels of interaction with the internet.

II. PRIVACY SHIELDING METHODOLOGY OVERVIEW

The process involves protecting confidential information of the users at four broad levels:

- *Shielding offline data*: This involves protection of the data that was stored on the computer system while accessing the internet and which is now saved for offline references. Such data are accessible without the internet and can also refer to the data that was never online.
- *Shielding data transmissions:* This involves protection of all the data that is being shared over the network with another party across the internet. It includes all communication mediums and transmission formats that occur between two users on the internet.
- *Shielding data stored by third party applications:* Third parties like certain software, websites, ISPs [1] etcetera collect sensitive information about the user and analyses certain patterns in one's usage. This data is stored and used as and how the application provider pleases, to improve the quality of the service. Also, there exists the data that the user chooses to agree for the application to store, such as voicemails or emails. This level deals with shielding against such consequences.
- *Shielding browsing data*: The browsing history and the search queries are stored by search engine providers in order to improve their algorithms and to reduce the time taken to fetch our search results with maximum accuracy. Such browsing data, can be confidential and if revealed would lead to a privacy breach. This level involves shielding against such surveillance.

All the above mentioned levels of interaction with the internet are readily exploited by government agencies with their limitless infrastructure and political might.

A systematic approach to each level is hence followed as part of the Privacy Shielding methodology to thwart any attempts of breaching one's privacy.

III. PRIVACY SHIELDING IMPLEMENTATION

A. *SHIELDING OFFLINE DATA*

---

[1] Internet Service Provider

1) *Secure File Encryptions*
- Encrypt Files Inside The Folders Using TRUECRYPT.

Reasons for choosing this software:

a. TRUECRYPT provides more evidence of honest functioning than any CPU maker, and RAM or flash memory manufacturer, or any computer retailer.

b. TRUECRYPT is open source and the source code is available for inspection by anyone at any time.

c. You can compile a binary from the source code and compare it to precompiled binaries.

d. Unintentional security flaws have been discovered in the source code and fixed by independent security professionals.

Main Features [5]:
a. Creates a virtual encrypted disk within a file and mounts it as a real disk.
b. Encrypts an entire partition or storage device such as USB flash drive or hard drive.
c. Encrypts a partition or drive where Windows is installed (pre-boot authentication).
d. Encryption is automatic, real-time (on-the-fly) and transparent.
e. Parallelization and pipelining allow data to be read and written as fast as if the drive was not encrypted.
f. Encryption can be hardware-accelerated on modern processors
g. Provides plausible deniability, in case an adversary forces you to reveal the password: Hidden volume (steganography) and hidden operating system.
Link: www.truecrypt.org

- Use Windows BitLocker drive encryption for drives with confidential files. Without it, anyone with a few minutes physical access to your computer, tablet or smartphone can copy its contents, even if they don't have your password.
Link:http://windows.microsoft.com/en-us/windows7/products/features/bitlocker

2) *Secure File System Conversions:* If file systems have not been updated, it is highly recommended to upgrade to NTFS file system. NTFS is the recommended file system because it is more powerful than FAT or FAT32, and includes features required for hosting *Active Directory* [2] as well as other important security features. You can use features such as Active Directory and domain–based security only by choosing NTFS as your file system.
Link: http://technet.microsoft.com/en-us/library/bb456984.aspx

3) *Secure File Deletions:* Discard confidential digital files safely using an effective file shredding software. Any deleted file could be recovered until it is being overwritten. Files in the recycle bin are not overwritten until you empty it. Based on the functionality, it is advised to choose from:
a. *DBAN*-Darik's Boot and Nuke is a boot disk that that securely deletes any contents of any hard disk. It is ideal for bulk or emergency data destruction as it can automatically wipe away the contents of any detected hard disk without a trace. This is the ultimate shredder.
Link: http://www.dban.org/download
b. *File Shredder*-This is a reliable source of secure file deletion that will completely erase files from your hard drive. It has a user-friendly interface. It has the functionality to let you browse and choose which files you want deleted. You can also choose 5 options for shredding algorithms.
Link: http://www.fileshredder.org/

B. SHIELDING DATA TRANSMISSIONS

---

[2] **Active Directory** (**AD**) is a directory service implemented by Microsoft for Windows domain networks. It is included in most Windows Server operating systems.

1) Using the TOR to connect to the internet: TOR [6], previously an acronym for The Onion Router [3] is free software for enabling online anonymity and censorship resistance. Tor directs Internet traffic through a free, worldwide, volunteer network consisting of more than five thousand relays to conceal a user's location or usage from anyone conducting network surveillance or traffic analysis. Using TOR makes it more difficult to trace Internet activity, including "visits to Web sites, online posts, instant messages, and other communication forms", back to the user and is intended to protect the personal privacy of users, as well as their freedom and ability to conduct confidential business by keeping their internet activities from being monitored.

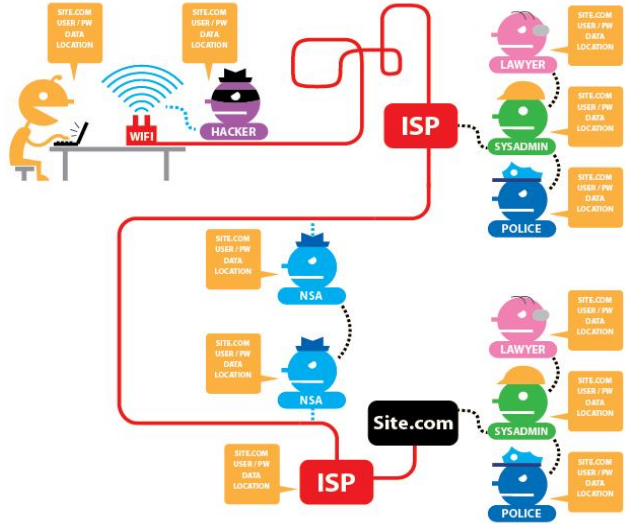

Fig. 1  Illustration of a network without TOR and https

The above is an illustration of the privacy scenario with respect to various actors in a network when the transmissions are carried out without the use of TOR and http secure mode.

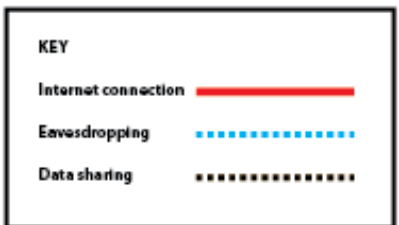

Fig. 2 Key representing the connections

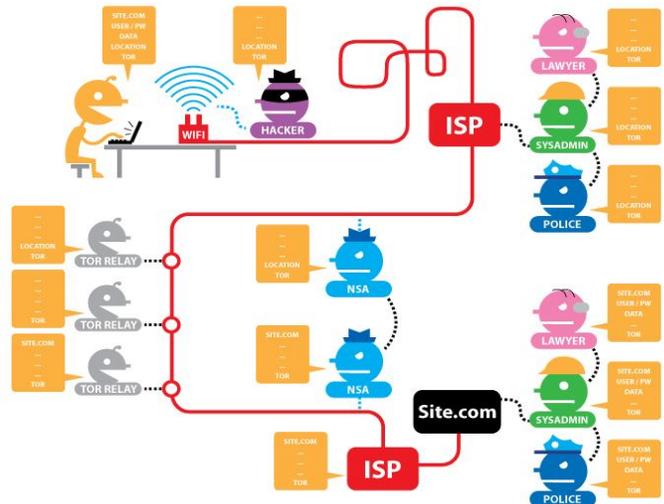

Fig. 3 Illustration of a network with TOR and https

Apart from TOR, many other separate forms of privacy shielding techniques can be incorporated during all data transmissions from the host system to the network. This can be used along with tor browsing to provide an added layer of security.

2) Protection for Connections to Web Sites: Use 'HTTPS Everywhere': By using an add-on to enforce https for Chrome or Firefox, one can maximise the amount of web data you protect by forcing websites to encrypt webpages whenever possible. HTTPS encrypts pages, and attempts to ensure three things: (1) that third parties cannot see the contents of the page; (2) that the page cannot be modified by third parties; (3) that the page was really sent by the web server listed in the URL bar [7].

3) Use encrypted mail like Hush mail: Hush mail uses industry standard algorithms as specified by the Open PGP Standard [4](RFC 2440) to ensure the security, privacy and authenticity of your email. We also protect all webmail traffic with HTTPS. With Hush mail, all you need to remember is your passphrase. Hush mail takes care of everything else in the background. This seamless and transparent encryption process makes Hush mail one of the most user-friendly secure email solutions available [8].

---

[3] **Onion Routing** refers to layers of encryption, nested like the layers of an onion, used to anonymize communication. TOR encrypts the original data, including the destination IP address, multiple times and sends it through a virtual circuit comprising successive, randomly selected TOR relays.

[4] **Pretty Good Privacy** (**PGP**) is a data encryption and decryption computer program that provides cryptographic privacy and authentication for data communication. PGP is often used for signing, encrypting, and decrypting texts, e-mails, files, and directories to increase the security of e-mail communications.

4) Use encrypted instant messaging (Off the Record): Off-the-Record (OTR) is a layer of encryption that can be added to any existing instant message chat system, provided that you can connect to that chat system using a chat client that supports OTR. With OTR it is possible to have secure, end-to-end encrypted conversations over services like Google Talk and Facebook chat without Google or Facebook ever having access to the contents of the conversations. Note: this is different than the "off-the-record" option in Google, which is **NOT** secure. Also, while Google and Facebook's HTTPS connection is very valuable for protection against your message while it's in transit, they still have the keys to your conversations so they can hand them over to authorities [9]. Some other instant messaging applications with high grade encryption that are also easy to use are:
- Chat Crypt – for websites and browser instant messaging
- Chat Secure (previously known as Gibber bot)- for android phones
- Crypto Cat

Link: http://www.chatcrypt.com/
Link: https://guardianproject.info/apps/chatsecure/
Link: https://crypto.cat/

5) Use encrypted VoIP communications: To prevent tapping and snooping on your phone calls by organizations, it is suggested to use encryption for VoIP as well. One open source secure application to achieve this is Silent Phone [9].
Link: https://silentcircle.com/web/silent-phone/

6) Use encrypted phone calls: Encryption can also be provided for phone calls over the network provider when a common application is being used by both the communicating parties. One such application is Red Phone. This application easily blends in with the native phone calling application to provide the same calling experience with encrypted communication.
Link: https://whispersystems.org/

7) Emphasize for Host-Proof Hosting: This refers to the service where the privacy of the client cannot even be breached by the host. This can be achieved in two ways: (1)Encrypting all the files using the file encryption methods described in the A1)Secure File encryptions section before uploading to the cloud server, and by (2) Using services that already provide Host-Proof Hosting to the clients. Such services are:
- Spider Oak
- Tar Snap
- Wuala

Link: https://spideroak.com/
Link: https://www.tarsnap.com/
Link: https://www.wuala.com/

## C. SHIELDING DATA STORED BY THIRD PARTY APPLICATIONS

Many third party applications we use today like Mail Providers, Internet Service Providers, Search Providers, Cloud Service Providers etcetera collect all sorts of information about us. The information can include:

- Name
- Address
- The length of time we have used that phone or Internet company
- Phone records including telephone number
- Records identifying all the phone numbers we have called or have called us, and the time and length of each call
- Internet records like the times we have signed on and off of the service, the length of each session, and the IP address that the ISP has assigned for each session
- Information on how we pay your bill, including any credit card or bank account number
- Location history with timeframes linking one's location in each timeframe to a map
- Transaction history of online purchases
- A comprehensive profile of us with respect to our interests based on searches and some keywords in the mail to provide personalized ads. This is called targeted marketing

There are ways to shield oneself from such sorts of surveillance by third parties and this requires a broader perspective of privacy along with the use of the privacy shielding techniques discussed in the previous sections.

Firstly, it is important for one to closely weigh certain factors before posting any information on the internet that could connect directly to oneself. Also we must be aware that not many third parties exist that has the legal immunity to resist any legal requests by the government or any court. So we must realize, if some third party has information collected about us at any point of time, it is accessible by the governments or legal organizations that are pursuing mass surveillance programs. Keeping in mind these realities, here are some factors we need to consider for shielding our privacy:

Before posting anything on the internet via the various platforms such as:

- Social networking media-Facebook, Google+, Twitter, Flickr, Instagram, LinkedIn, Quora, Foursquare etcetera.

- Blogs- WordPress, Blogger, Tumblr, Medium, Squarespace etcetera.
- Forums
- Chat rooms
- IRC[5] chats,

it is important to ask some questions to oneself before thinking to post anything openly on the net, "Is it really necessary for me to post this particular opinion/ view/ information about me or the people I am connected to be openly visible on the internet?"

"Do I really want the phone company to have a record of this call-who I called, when, and how long I talked?"

"Do I really want a copy of this email floating around in the recipient's inbox, or on my email provider's system?"

"Do I really want my cell phone provider to have a copy of that embarrassing SMS text message?"

"Do I really want Google to know what I'm searching for?"

The answers to these questions can save sufficient effort in shielding one's privacy. However, if one feels offended to hold oneself back from freely expressing views and feels the 'Freedom of Speech and Expression' is being breached, then the techniques mentioned in the previous section B. Shielding Data Transmissions, can help. Also some changes to the settings of third party applications can help. Some such settings can include:

- Reviewing privacy settings of social networking sites and customizing them to suit privacy needs. Settings for Facebook and Google+ are shown:

Facebook Settings [10]

To view and adjust your privacy settings:

- Click  in the upper-right corner of any Facebook page
- Select **Settings** from the dropdown menu
- Select **Privacy** on the left
- Click on a setting (ex: **Who can see your future posts?**) to edit it

Selecting an Audience for Stuff You Share

You'll find an audience selector tool most places you share status updates, photos and other stuff. Just click the tool and select who you want to share something with.

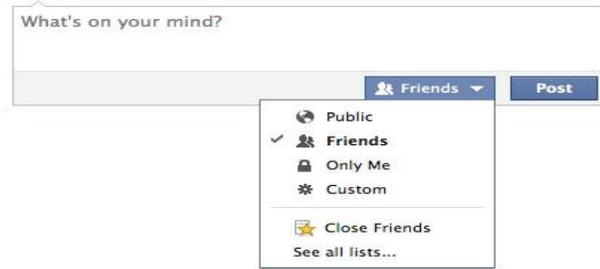

Fig. 4 Facebook audience selection for posts shared

How do I control who can see what's on my Timeline?

You can share basic information like your hometown or birthday when you edit your Timeline.
- Click **Update Info** (under your cover photo).
- Then click the **Edit** button next to the box you want to edit. Use the audience selector next to each piece of information to choose who can see that info.
- Anyone can see your public information, which includes your name, profile picture, cover photo, gender, username, user ID (account number), and networks.

Choose who to share information with

- Sign in to Google+. Place your cursor in the top left corner of the Google+ main menu and click 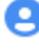 **Profile.**
- Click the **About** tab below the bottom left corner of your cover photo.
- You'll see boxes containing different types of information you can add. Click the **Edit** link at the bottom of the box you want to adjust.
- You can change who sees the information in the box using a dropdown menu in the upper-right corner of the box. You can choose to share the information with: "Extended circles," "Public," "Your circles," or "Only you"

Choose which of your Profile tabs are visible

When viewing a profile, tabs are located at the top of the page. The "About" tab and "Posts" tab will always appear to anyone who visits the profile page. You may have other tabs on your profile like "Photos," "Videos," "+1" and "Reviews". You can use your settings to select which tabs are visible to people.

---

[5] Internet Relay Chat

Google+ settings [11]

- Place your cursor in the top left corner for the Google+ main menu and click **Settings**.
- In the **Profile** section, check or uncheck the tabs you'd like people to be able to see.

▪ Verify the settings of any such other platforms which is used such as blogs and forums and customize to improve privacy.

▪ Shield location privacy- Many third party applications store your location by means of your IP address. Google, for instance, logs your entire location history against each time frame of the day from the devices you use. Google location history can be checked from the link: https://maps.google.com/locationhistory

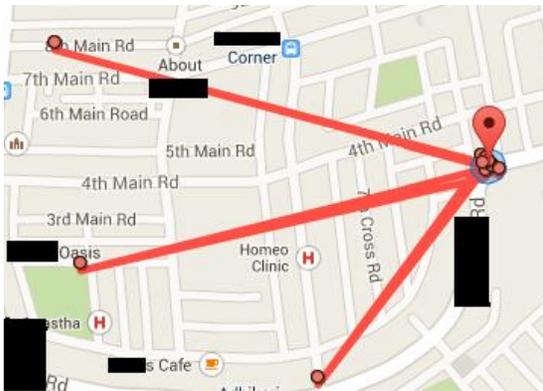

Fig. 5 Google Location History on a particular date

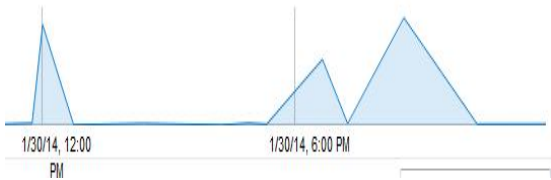

Fig. 6 Timeline graph for Google Locations for the date

Above is the censored version of a google location result for a google user. Access to this information from google to any vested interest can prove to be very dangerous for the victim and is a serious breach of privacy.

Change the google privacy settings to stop google from collecting any location information about you. This can be done from any device.

- Go to Google Settings
- Click on Location
- Un-tick **Access Location**
- Change **Location Reporting** to OFF
- Change **Location History** to OFF

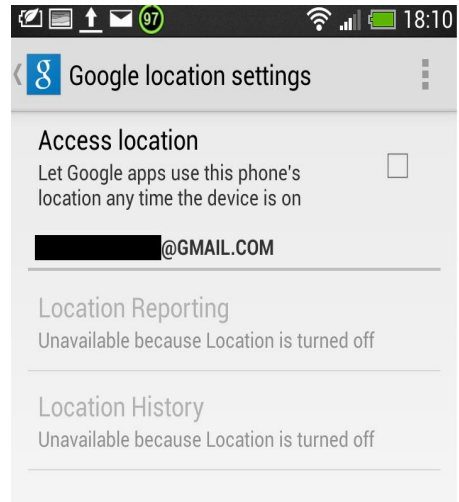

Fig. 7 Google Location Settings for a device

Also change the settings from your Google+ profile to prevent it from collecting location data.

Fig. 8 Google location settings for Google+ profile

For shielding location privacy, google is not the only source. It is advised to carefully read the 'App Permissions' section for a better picture before using any application for any device

▪ Speaking of shielding privacy from third party applications, it is also possible that some parts of our digital identity can be stealthily shared to the internet by the operating system we use. For example, Microsoft Windows Update Centre sends information to the Microsoft servers on a timely basis which in turn can also have some information of the user. This begs the need for even a more secure Operating System for paranoid users to protect one's identity and remain anonymous. **The Amnesic Incognito Live System** or **Tails** [12] is designed just to achieve that. Tails is a security-focused Debian-based Linux distribution aimed at preserving privacy and anonymity It is the next iteration of development on the previous Gentoo-based Incognito Linux distribution. All its outgoing connections are forced to go through TOR and direct (non-anonymous) connections are blocked. The system is designed to be booted as a live DVD or live USB, and will leave no trace (digital footprint) on the machine unless explicitly told to do so.

*D.  SHIELDING BROWSING DATA*

This level deals mainly with shielding search privacy and the information that can be obtained from browsing the internet.

Shielding search privacy requires the following steps:
- Use anonymous search engines like Duck Duck Go search.
  Link: https://duckduckgo.com/
- Do not search for any personal information like one's own name, email id, address etcetera.
- Set search settings to not store any search information in the browser.
- Do not allow browser to store cookies by changing the settings in the browser. It is advised to use Mozilla Firefox for such settings as it is open source and is less likely to have backdoors in the source code to support any surveillance programs:

  - From the "**Edit**" menu, select "**Preferences**"
  - Click on "**Privacy**"
  - Select the "Cookies" tab
  - Set "Keep Cookies" to "until I close Firefox"
  - Click on "Exceptions," type in the domains of all of your search sites, and choose "Block" for all of them

- Do not log in to the browser or the search engine.
- Do not use the search engine provided by the ISP.

Shielding browsing data requires the following precautions:
- For anonymous browsing it is good to use VPNs or TOR networks mentioned in B 1)
- Browser add-ons [13] do a great deal of help in shielding browser data:

  - No Script Security Suite- allows active content to run from only trusted sites, and protects from XSS and Click Jacking attacks.
  - Ghostery- that prevents websites from collecting data about the users.
  - No Trace- to limit the diffusion of personally identifiable information and protect user's privacy against Web tracking and other dangerous privacy threats.
  - Ad block plus to prevent unwanted ads from showing on the screen.
  - Ad block plus Pop up block add-extends the blocking functionality to Ad block plus to those annoying pop-up windows that open on mouse clicks and other gestures.
  - JavaScript De-obfuscator - This add-on will show you what JavaScript gets to run on a web page, even if it is obfuscated and generated on the fly. Simply open JavaScript De-obfuscator from the Tools menu and watch the scripts being compiled/ executed.

- Also change browser settings to suit privacy and security needs. Again, Mozilla Firefox is advised:
  - In Mozilla Firefox, click on the **Firefox** tab on the top left.
  - In the drop down menu, click **Options**.
  - Under the **Privacy** tab, check on the "**Tell sites I do not want to be tracked**" option and check on "**Clear history when Firefox closes**" option.
  - Under the **Security** tab, check on "**block reported attack sites**" and "**block report web forgeries**" options.

IV. CONCLUSIONS

This paper, 'Privacy Shielding against Mass Surveillance' is a humble attempt in providing comprehensive guidelines for internet users to protect one's identity while browsing freely. Although maximum effort has been put in to provide with the most effective strategies supporting the objective, this is not a fool proof guide to complete privacy. With change in time, technology, privacy laws and circumstances change. It is important for one to be constantly updated on the latest advancements and to make a conscious effort to keep up with the technology and attempt to maintain privacy. As mentioned earlier, Internet of Things will be the next big thing and this presses the need for self-awareness in the field of information security. All the techniques mentioned in this paper will serve as a foundation stone for a long journey in shielding privacy as technology progresses forward.

Privacy is a virtue and with current surveillance trends, shielding privacy better be a way a life!